\documentclass[lettersize,journal]{IEEEtran}

\usepackage{hyperref}
\usepackage{cleveref}
\usepackage{multirow}
\usepackage{siunitx}
\usepackage{subcaption}
\usepackage[labelfont=bf]{caption}
\usepackage{tabu}
\usepackage[table,xcdraw]{xcolor}
\usepackage{textgreek}
\usepackage{todonotes}
\presetkeys{todonotes}{}{}
\usepackage[commandnameprefix=ifneeded]{changes}
\usepackage{xfp}
\usepackage{xspace}
\usepackage{colortbl}

\usepackage{tikz}
\usepackage{calc}
\usepackage{pgfplots}
\usepackage{pgfplotstable}
\usetikzlibrary{fit, calc}
\usepgfplotslibrary{colorbrewer}

\usepackage{booktabs}  

\newcommand{\eg}{e.g.}

\newcommand{\name}{Celeris\xspace}
\definecolor{CB-Pl-blue}{HTML}{a6cee3}
\definecolor{CB-Pd-blue}{HTML}{1F78b4}
\colorlet{blue-highlight}{CB-Pd-blue!75!black}
\definecolor{CB-Pl-green}{HTML}{b2dF8a}
\definecolor{CB-Pd-green}{HTML}{33a02c}
\definecolor{CB-Pl-red}{HTML}{Fb9a99}
\definecolor{CB-Pd-red}{HTML}{e31a1c}
\definecolor{CB-Pl-orange}{HTML}{FdbF6F}
\definecolor{CB-Pd-orange}{HTML}{FF7F00}
\definecolor{CB-Pl-purple}{HTML}{cab2d6}
\definecolor{CB-Pd-purple}{HTML}{6a3d9a}
\definecolor{CB-Pl-brown}{HTML}{FFFF99}
\definecolor{CB-Pd-brown}{HTML}{b15928}
\pgfplotscreateplotcyclelist{PairedLightDark}{%
  {CB-Pd-blue,draw=none},
  {CB-Pl-blue,draw=none},
  {CB-Pd-green,draw=none},
  {CB-Pl-green,draw=none}}
\pgfplotscreateplotcyclelist{PairedLightDarkLinesPoints}{%
  {CB-Pd-blue,solid,thick,mark options=solid,mark=*},
  {CB-Pd-green,densely dotted,very thick,mark options=solid,mark=*},
  {CB-Pd-red,densely dashdotted,thick,mark options=solid,mark=*},
  {CB-Pd-brown,densely dashed,thick,mark options=solid,mark=*},
  {CB-Pd-purple,dashdotted,thick,mark options=solid,mark=*},
  {CB-Pd-orange,loosely dashed,thick,mark options=solid,mark=*}}
\pgfplotscreateplotcyclelist{PairedLightDarkLines}{%
  {CB-Pd-blue,solid,thick},
  {CB-Pd-green,densely dotted,very thick},
  {CB-Pd-red,densely dashed,thick},
  {CB-Pd-brown,densely dotted,thick}}

\pgfplotsset{compat=newest}
  \pgfplotsset{
    g/.style={
        cycle list name=PairedLightDarkLinesPoints,
        width=\linewidth,
        height=1.25in,
        no marks,
        xlabel={Load (\%)},
        axis x line*=bottom,
        axis y line*=left,
          font=\footnotesize\sffamily,
        set layers,
        xmax=150,
        legend style={
          legend columns=4,
          at={(0.5,-0.5)},
          anchor=north,
          draw=none,
          /tikz/every even column/.append style={column sep=0mm},
        },
        }
      }
  \pgfplotsset{
    sloline/.style={
          mark=none,
          thick,
          black,
        }
      }
  \pgfplotsset{
    latplot/.style={
        g,
        ylabel={Latency (Norm.)},
        legend entries={50\%,90\%,99\%,99.9\%},
        ymin=0,
        ymax=4,
        }
      }
  \pgfplotsset{
    accplot/.style={
        g,
        ylabel={Accuracy (\%)},
        ymin=80,
        ymax=95,
      }
    }

\newlength{\zeroheight}

\makeatletter
\NewDocumentCommand {\getnodedimen} {O{\nodewidth} O{\nodeheight} m} {
  \begin{pgfinterruptboundingbox}
  \begin{scope}[local bounding box=bb@temp]
    \node[inner sep=0pt, fit=(#3)] {};
  \end{scope}
  \path ($(bb@temp.north east)-(bb@temp.south west)$);
  \end{pgfinterruptboundingbox}
  \pgfgetlastxy{#1}{#2}
}
\makeatother

\ifCLASSOPTIONcompsoc
  \usepackage[nocompress]{cite}
\else
  \usepackage{cite}
\fi

\ifCLASSINFOpdf
\else
\fi

\begin{document}
\title{Reimagining RDMA Through the Lens of ML \\ 
       \vspace{-5pt}}
\author{{Ertza Warraich, Ali Imran, Annus Zulfiqar, Shay Vargaftik, Sonia Fahmy, and Muhammad Shahbaz}
\vspace{-28pt}
}

\maketitle  

% \IEEEtitleabstractindextext{
\begin{abstract}
As distributed machine learning (ML) workloads scale to thousands of GPUs connected by ultra-high-speed interconnects, tail latency in collective communication has emerged as a primary bottleneck. 
Prior RDMA designs, like RoCE, IRN, and SRNIC, enforce strict reliability and in-order delivery, relying on retransmissions and packet sequencing to ensure correctness. 
While effective for general-purpose workloads, these mechanisms introduce complexity and latency that scale poorly, where even rare packet losses or delays can consistently degrade system performance.
We introduce \name{}, a domain-specific RDMA transport that revisits traditional reliability guarantees based on ML's tolerance for lost or partial data.
\name{} removes retransmissions and in-order delivery from the RDMA NIC, enabling best-effort transport that exploits the robustness of ML workloads.
It retains congestion control (\eg, DCQCN) and manages communication with software-level mechanisms such as adaptive timeouts and data prioritization, while shifting loss recovery to the ML pipeline (\eg, using the Hadamard Transform).
Early results show that \name{} reduces 99th-percentile latency by up to 2.3$\times$, cuts BRAM usage by 67\%, and nearly doubles NIC resilience to faults---delivering a resilient, scalable transport tailored for ML at cluster scale.
\vspace{-3pt}
\end{abstract}

% }

\thispagestyle{empty}
\pagestyle{empty} 

\vspace{-8pt}
\ifCLASSOPTIONcompsoc
\IEEEraisesectionheading{\section{Introduction}\label{sec:introduction}}
\else
\section{Introduction}
\label{sec:introduction}
\fi
\IEEEPARstart{D}{istributed} machine learning (ML) workloads (training and inference) now span thousands of GPUs connected by ultra-high-speed 100--400G interconnects. 
As models grow and clusters scale, the primary bottleneck shifts from compute to communication~\cite{warraich2025optireduce,wang2024towards}. 
Collective operations such as AllReduce, AllGather, and All-to-All lie on the critical path of both data- and model-parallel processing. 
These operations introduce global synchronization points, where even minor transport delays can stall progress across the entire cluster. 
As a result, tail latency---not average throughput---has become the dominant limiter of model efficiency at scale~\cite{warraich2025optireduce, wang2024towards}.

To mitigate communication overhead, the community has focused on optimizing collective algorithms (\eg, NCCL, RCCL, and MSCCL) and reducing traffic via sparsification, quantization, and other compression techniques~\cite{fei2021efficient, alistarh2017qsgd}. 
These advances build on a foundational insight: ML workloads processed with stochastic gradient descent (SGD) are statistically resilient; they tolerate noise, approximation, and even bounded loss without hurting convergence~\cite{warraich2025optireduce,wang2024towards}.

Yet, the underlying transport stack remains overly conservative (\Cref{tab:rdma-evolution}).
RoCE, the de facto RDMA protocol in ML clusters, is designed for strict reliability and correctness~\cite{gangidi2024rdma}.
It enforces in-order delivery, retransmissions, and lossless operation using Priority Flow Control (PFC). 
While effective for general-purpose workloads (\eg, key-value stores, distributed databases, and RPCs), these mechanisms increase hardware complexity and introduce latency spikes that scale poorly with cluster size. 
A single packet drop can trigger go-back-N retransmissions or cascade into fabric-wide PFC stalls, inflating tail latency and limiting scalability~\cite{wang2023srnic}.

Efforts like IRN~\cite{mittal2018revisiting} eliminate PFC by handling packet loss directly in the NIC.
IRN uses selective repeat with bitmap tracking and SACK-based recovery to improve scalability. 
However, its reliance on NIC-resident bitmaps and reordering logic increases per-QP state, placing pressure on NIC memory and constraining overall connection density. 
SRNIC~\cite{wang2023srnic} addresses this by offloading retransmission and reordering to host software, while eliminating WQE caching to further simplify NIC design.
Its hybrid approach assumes that packet loss is rare and delegates recovery to the software slow path.
However, at ML scale, that assumption no longer holds: what appears as rare losses at the single-node level becomes frequent when viewed across thousands of GPU nodes synchronizing in parallel.
These losses accumulate at synchronization points, exacerbated by slow workers, turning infrequent delays into persistent tail latency---a classic tail-at-scale effect~\cite{dean2013tail}.

This paper revisits the NIC transport design from a domain-specific perspective. We ask: {\em if ML can tolerate partial loss and reordering, why pay the cost of enforcing strict delivery guarantees at the RDMA NIC transport layer?}

To that goal, we make a case for \name{}, a hardware-accelerated RDMA transport tailored for ML workloads. 
\name{} removes retransmissions and in-order delivery, forwarding best-effort, unordered packets directly to application memory.
It retains congestion control (\eg, via DCQCN), while delegating timeout handling and data prioritization to software.
This architecture drastically simplifies NIC logic, reducing memory footprint and fault exposure.
Rather than recovering from packet loss within the transport layer, \name{} bounds its impact and leverages recovery mechanisms within ML pipelines itself---such as Hadamard Transform~\cite{warraich2025optireduce}.

By aligning transport semantics with ML workload characteristics, \name{} avoids the systemic buildup of tail latency from cluster-wide delays due to packet loss and stragglers. 
Preliminary results from our FPGA prototype and simulation-based evaluations show that \name{} reduces 99th-percentile latency by up to 2.3$\times$, cuts BRAM usage by 67\%, and nearly doubles hardware fault resilience. 
These early results suggest that rethinking transport guarantees through an ML-centric lens can unlock significant gains in performance, scalability, and resilience.

\begin{table*}[t]
\centering
\begin{tabular}{@{}l|l|l|l|>{\columncolor[HTML]{cfe2f3}}l}
\toprule
\textbf{Design Aspect}               & \textbf{RoCE} & \textbf{IRN}~\cite{mittal2018revisiting} & \textbf{SRNIC}~\cite{wang2023srnic} & \textbf{\name{}} \\ \midrule
PFC Required                        & Yes                 & No                        & No                          & No \\
Transport Reliability~~~~               & Go-back-N~~~~~~~~~~           & Selective Repeat~~~~~~          & Selective Repeat (SW)~~~~ & None (best-effort)~~~~~~~~ \\
Packet Reordering                   & Dropped             & Buffered in NIC           & SW reordering             & Offset-based placement \\
Congestion Control                    & Hardware               & Hardware                     & Hardware             & Hardware \\
WQE Cache                           & Present (HW)        & Present (HW)              & Eliminated              & Eliminated \\
NIC State per QP                    & 407B               & 596B                     & 242B                    & 52B \\
QP Scalability                      & 10K                 & 8K                        & 20K                         & 80K \\
Target Workloads                    & General RDMA        & General RDMA              & General RDMA + ML             & ML Collectives  \\ \midrule
\textbf{Core Focus}                 & High performance    & +Network efficient      & +Connection scalable         & +Tail optimal \\ \bottomrule
\end{tabular}
\vspace{-4pt}
\caption{\bf Comparison of RDMA NIC designs: \name{} unifies performance, efficiency, and scalability from prior work, while adding tail-optimized support for loss-tolerant ML collectives.}
\label{tab:rdma-evolution}
\vspace{-18pt}
\end{table*}

\vspace{-11pt}
\section{Background and Motivation}
\label{sec:background}
%\subsection{Communication Bottlenecks in ML Workloads}
\noindent{\em A. Communication Bottlenecks in ML Workloads
\vspace{3pt}\\}
Modern ML workloads---both training and inference---depend heavily on collective communication across large clusters of accelerators. 
These collectives enable a variety of parallelism strategies: 
Data parallelism aggregates gradients using AllReduce. 
Tensor and pipeline parallelism, employed in both training and inference, exchange activations and weight shards using ReduceScatter/AllGather (and occasionally AllReduce)~\cite{rajbhandari2022deepspeed}. 
For long-sequence inference, context parallelism partitions the attention context and performs step-wise All-to-All exchanges (for K/V or token blocks) along with localized AllGather/ReduceScatter for residuals and projections. 
% \ertza{Response point 6}
% \textcolor{blue}{
Expert parallelism (Mixture-of-Experts, MoE) uses All-to-All communication to dynamically route tokens between experts. 
And, hybrid parallelism blends these strategies to support increasingly complex workloads.
% }

\begin{figure}[t]
  \centering
  \begin{subfigure}[b]{0.39\linewidth}
    \centering
    \includegraphics[width=\textwidth]{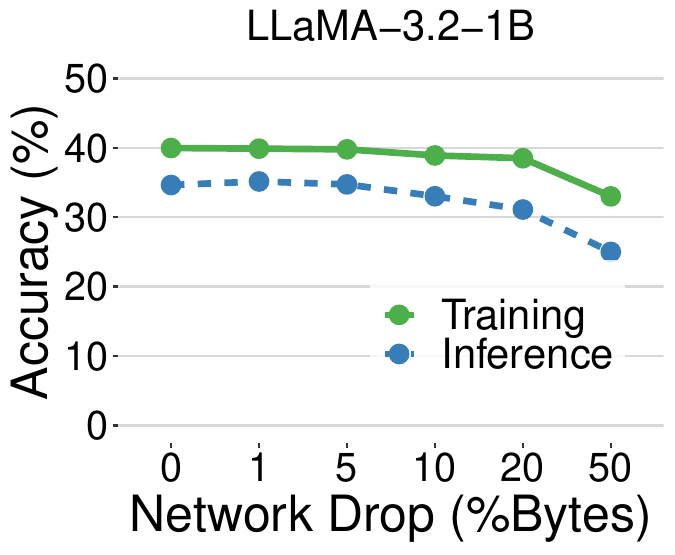}
    \vspace{-15pt}
    \caption{Training and inference on the ARC dataset.}
    \label{fig:training}
  \end{subfigure}
  \hfill
  \begin{subfigure}[b]{0.59\linewidth}
    \centering
    \includegraphics[width=\textwidth]{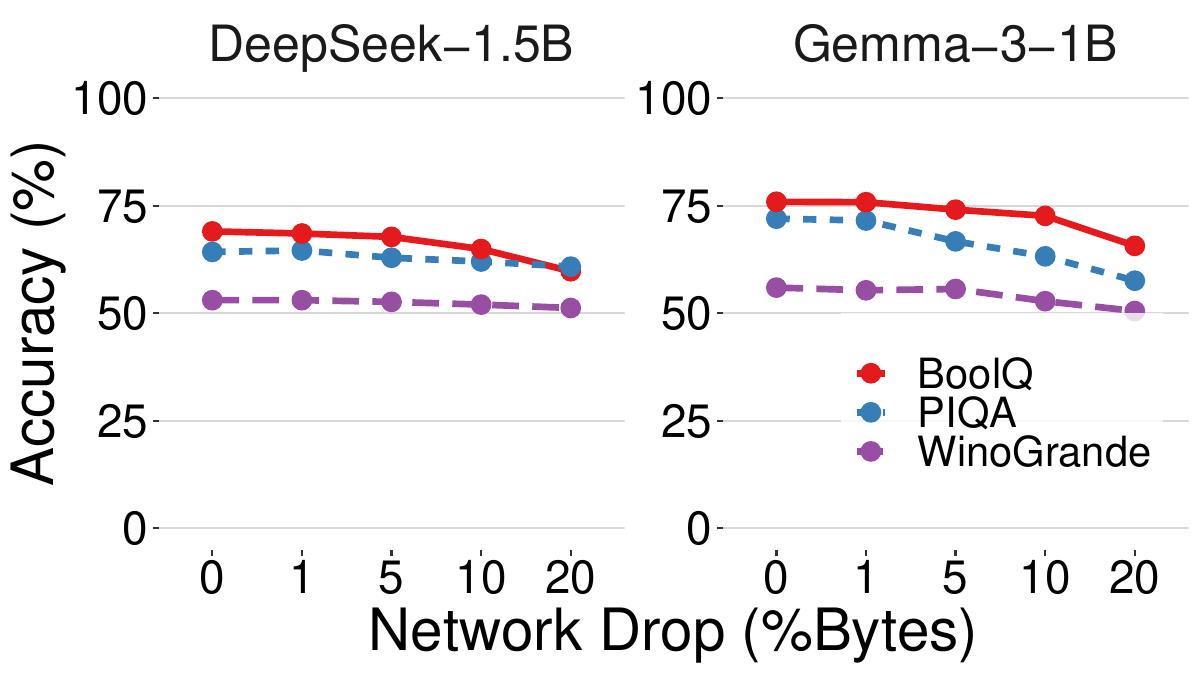}
	\vspace{-15pt}
    \caption{Inference-only accuracy using multiple datasets (BoolQ, PIQA, and more).}
    \label{fig:inference}
  \end{subfigure}
  \vspace{-15pt}
  \caption{\bf Training and inference accuracy of all models remains stable under partial network drops ($\leq$ 5\%).}
  \label{fig:loss-resilience}
  \vspace{-18pt}
\end{figure}

As model sizes and cluster scales grow, these collective operations increasingly dominate end-to-end performance~\cite{gangidi2024rdma, warraich2025optireduce}. 
Because they synchronize large numbers of accelerators, even rare stragglers can stall an entire iteration, and their overhead grows rapidly with system scale. 
Yet the data they exchange---intermediate tensors, activations, routing metadata, and partial outputs---are often transient, later aggregated, recomputed, or subsampled downstream. 
This creates a fundamental mismatch: today's transports enforce strict in-order reliable delivery, while what matters most to ML workloads is minimizing collective tail latency. 
Empirical studies show that collectives can consume up to 70\% of iteration time in large-scale systems~\cite{warraich2025optireduce}, highlighting their role as the dominant communication bottleneck.

\vspace{3pt}
\noindent{\em B. ML is Resilient to Loss
\vspace{3pt}\\}
Despite their sensitivity to tail latency, ML workloads are inherently robust to packet loss and partial delivery. 
Stochastic gradient descent (SGD) naturally smooths over noise and missing updates, and recent work has exploited this tolerance to reduce bandwidth via gradient sparsification, quantization, and \textsf{bfloat16} compression~\cite{alistarh2017qsgd, fei2021efficient}. 
In-network aggregation systems, such as NVIDIA SHARP, perform lossy reductions in the dataplane without hurting convergence.
And as shown in \cite{warraich2025optireduce, wang2024towards}, models maintain high accuracy even under substantial gradient drops during training communication (\Cref{fig:loss-resilience}a).

Importantly, this resilience is not limited to gradients. 
Intermediate tensors, activations, and routing metadata---often exchanged in collectives like AllGather and All-to-All---are frequently sparse, recomputable, or overwritten in later iterations. 
Even in inference~\cite{rajbhandari2022deepspeed}, partial loss can be absorbed through redundancy, expert fallback paths, or statistical softmax smoothing (\Cref{fig:loss-resilience}b). 
This tolerance opens the door to a radically simpler transport: one that delivers timely data, even if it is occasionally incomplete.\footnote{
% \textcolor{blue}{
Best-effort transport introduces nondeterminism, but this is acceptable in large-scale LLMs~\cite{warraich2025optireduce,wang2024towards} and can be mitigated through per-step logging of lost data to reproduce outcomes during debugging.}
% }
% \shahbaz{add citation}
%\ertza{Response point 4}
%\textcolor{orange}{
%A known limitation of using best-effort transport is reduced reproducibility, as transient loss can make runs nondeterministic. However, this trade-off is acceptable in large-scale LLM workloads, where nondeterminism is already inherent, even models configured for determinism (fixed seeds, zero sampling) exhibit variance in accuracy and outputs~\cite{atil2024non}. Additionally, per-step logging of lost data (missing bytes and offsets) can further aid debugging and improve reproducibility.
%}

%\vspace{-10pt}
%\subsection{The Cost of Reliable Transports}
\vspace{3pt}
\noindent{\em C. The Cost of Reliable Transports
\vspace{3pt}\\}
Mainstream RDMA transports like RoCE enforce strict delivery guarantees---go-back-N retransmissions, in-order delivery, and loss prevention via Priority Flow Control (PFC). 
These features ensure correctness but require deep integration of control state into the NIC: reorder queues, sequence number tables, retry logic, and recovery timers are tightly coupled with the datapath. 
While effective for storage and database workloads, this design scales poorly in large ML clusters, where collective synchronization amplifies even rare packet drops into full-step stalls.

Efforts like IRN~\cite{mittal2018revisiting} remove PFC and replace go-back-N with selective repeat and SACK-based recovery, improving network scalability. 
SRNIC~\cite{wang2023srnic} goes further by offloading reordering and retransmissions to software and removing the WQE cache. 
These designs reduce NIC memory usage (Table~\ref{tab:rdma-evolution}), but still retain transport-layer recovery. 
As a result, they continue to trigger slow-path handling under loss---a mechanism that becomes problematic at scale. 
In large jobs, what appears rare at a single node becomes frequent in aggregate, creating persistent tail latency through tail-at-scale effects~\cite{dean2013tail}.
Meta has demonstrated limited success with RoCE in carefully tuned AI clusters~\cite{gangidi2024rdma}, but only under tight (HPC-like) conditions---dedicated workloads, fixed topologies, and specialized routing---conditions infeasible in typical datacenter or cloud environments.

%\vspace{-8pt}
%\subsection{Reliability Hurts Resilience}
\vspace{3pt}
\noindent{\em D. Reliability Hurts Resilience
\vspace{3pt}\\}
Beyond latency, the real cost of enforcing transport reliability is reduced system resilience. 
Per-QP state---retry counters, sequence numbers, recovery timers---is typically stored in on-chip SRAM. 
These structures are large, tightly coupled with critical datapath logic, and vulnerable to soft errors~\cite{amd-seu-estimator}. 
Even if each NIC reports a high Mean Time Between Failures (MTBF), \eg, 400,000 hours, at a 10,000-node scale, faults can occur every 40 hours. 
Worse, these faults often affect precisely the components responsible for enforcing correctness: a stuck retry timer or corrupted sequence number can silently stall a QP and block an entire collective~\cite{warraich2025optireduce, wang2024towards}.

Our key observation is that ML workloads do not require these mechanisms to make progress or maintain correctness. 
Instead of hardening unreliable machinery, we eliminate it. 
By removing retransmissions, reordering, and per-packet tracking entirely, \name{} (\S\ref{sec:design}) reduces per-QP state to just 20 bytes. 
There are no retry counters, timers, or window logic---only minimal metadata required to push data. 
This not only improves tail latency but nearly doubles hardware fault tolerance, as shown in our FPGA prototype and MTBF analysis (\S\ref{sec:evaluation}).

\vspace{-6pt}
\section{\name{}: A Tail-Optimal RDMA NIC}
\label{sec:design}

%\ertza{Response point 5}
%\textcolor{orange}{
%While OptiReduce~\cite{warraich2025optireduce} and MLT~\cite{wang2024towards} introduced loss-tolerant communication for data-parallel training, \name{} generalizes this concept in two key directions. 
%First, it extends bounded-loss communication beyond gradient aggregation to all major ML communication patterns---including tensor, pipeline, context, and expert parallelism---where the exchanged data (parameters, activations, or tokens) cannot simply be overwritten in the next iteration. This broader scope requires tighter control on loss-bound, and mechanisms to preserve convergence. 
%Second, \name{} realizes this design inside RDMA hardware, where reliability, flow control, and completion semantics are deeply intertwined. Implementing a best-effort transport in this setting introduces unique challenges, \eg, the NIC must sustain congestion control fairness with no retransmission or reorder state, similarly the software coordinates synchronized timeouts and signals the hardware to stop the communication when the allocated time-window expires.}
% \textcolor{blue}{
\name{} is a domain-specific RDMA transport that removes traditional delivery guarantees to prioritize simplicity, scalability, and tail-optimal performance.
Unlike prior work that implements loss-tolerant communication in software (\eg, OptiReduce~\cite{warraich2025optireduce} and MLT~\cite{wang2024towards}), \name{} extends this approach through a co-designed hardware-software implementation.
% }
Instead of enforcing retransmissions or in-order delivery, \name{} delivers best-effort, unordered data directly to application memory---exploiting the statistical resilience of ML workloads (\S\ref{sec:background}).
It minimizes NIC state, bounds communication via software-driven timeouts, and shifts loss handling to the ML framework.
This design reduces tail latency and hardware overhead, enabling efficient scaling across thousands of GPUs without compromising model convergence.

%\vspace{-12pt}
%\subsection{Hardware -- Stateless RDMA Transport}
%\label{ssec:rdma-hw}
\vspace{4pt}
\noindent{\em A. Hardware -- Stateless RDMA Transport
\vspace{3pt}\\}
\name{} retains the foundational structure of conventional RDMA NICs: it supports QP-based communication, uses DMA to move data from GPU memory, and implements flow-level congestion control (\eg, DCQCN).
However, it departs from prior designs like RoCE and IRN~\cite{mittal2018revisiting} by eliminating all packet-level reliability mechanisms.
There are no retransmissions, reorder queues, selective repeat buffers, or outstanding request tables in the NIC.

Instead, each packet includes a logical offset into the target buffer, allowing direct placement without requiring reassembly or sequencing~\cite{mittal2018revisiting,wang2023srnic}.
This enables unordered delivery: packets are placed as they arrive, with no NIC-side tracking of order or completion.
The NIC is unaware of loss or duplication---it simply forwards data.

With no per-packet state or bookkeeping, the per-QP context is dramatically reduced---only 20 bytes (plus 32 bytes in case of DCQCN), compared to hundreds of bytes in RoCE, IRN, and SRNIC (Table~\ref{tab:rdma-evolution}).
This compact context allows the NIC to support 10$\times$ more connections using the same SRAM budget, enhancing scalability across large ML clusters.
The only active datapath logic is congestion control, which remains in hardware to ensure fair bandwidth allocation.\footnote{
% \textcolor{blue}{
A software variant atop RDMA's UC and UD is possible, but would shift congestion control to software, increasing CPU overhead and sacrificing \name{}'s hardware simplicity.}
% }
%\ertza{Response point 2}
%\textcolor{orange}{
%% UCCL~\cite{zhou2025extensible} builds atop RDMA’s Unreliable Connection (UC) transport to operate with existing NICs, moving reliability and congestion control to software but making these decisions at a coarse \SI{32}{KB} chunk granularity to reduce the CPU overhead of per-packet control. 
%A variant of \name{} could be implemented on top of RDMA's Unreliable Connection (UC) or Unreliable Datagram (UD) for easier deployment with existing infrastructure. However, UC and UD do not implement any congestion control, as well as drop all out-of-order packets, leading to excessive drops where a single congestion event or an early packet loss can cause the remainder of message to be discarded. Dividing the message into smaller chunks and implementing a congestion control in software can somewhat reduce this loss, but increase the control overhead. Such a design would also forfeit \name{}’s advantage of reduced hardware state and logic, though it would still retain some tail-latency benefits.
%}

\name{} removes all reliability mechanisms from the NIC, and simplifies the datapath to a streamlined push engine focused purely on data movement.
This reduces BRAM usage, eliminates fault-prone state, and enables predictable, low-latency communication at scale.
The design aligns with a core insight of \name{}: if ML workloads can tolerate loss, enforcing strict delivery guarantees in hardware is not only unnecessary, it is counterproductive.

%\vspace{-12pt}
%\subsection{Software -- Bounded Timeouts \& Cluster Coordination}
%\label{ssec:rdma-sw}
\vspace{4pt}
\noindent{\em B. Software -- Bounded Timeouts \& Cluster Coordination
\vspace{3pt}\\}
In \name{}, the software stack assumes responsibility for progress and coordination, replacing NIC-managed reliability with bounded delivery windows and lightweight recovery strategies.
Each collective operation begins with the sender issuing WQEs to transmit data.
The NIC streams packets toward receivers without tracking their delivery or acknowledgment.

% \textcolor{blue}{
On the receiver side, software defines a step-level timeout that bounds the delivery window for each collective operation.
Packets arriving after this window are discarded.
At the end of the timeout, the collective step is finalized using only the data that arrived on time.
Because distributed training typically involves multiple concurrent collectives, such as data, tensor, or expert-parallel groups, the software maintains an independent timeout profile for each.
% } 
%%\ertza{Response point 3}
%\textcolor{orange}{
%Because distributed training typically involves multiple concurrent collectives---such as data, tensor, or expert-parallel groups---the software maintains an independent timeout profile for each. The collective layer, which orchestrates these operations, selects the appropriate timeout based on the active collective.
%}

Timeouts are dynamically adjusted.
After each step, the system measures how much data was received and how long the step took.
If all data arrived, the next timeout is updated to match the observed duration.
If only partial data was received, the system estimates the required duration for full delivery and sets the next timeout accordingly~\cite{warraich2025optireduce}.
These updates are smoothed using exponential averaging and bounded within a fixed range to ensure stable behavior.

To maintain cluster-wide coordination, nodes share their local timeout estimates at the end of each step.
All nodes then use the median of the reported values for the next round.
This synchronization prevents stragglers from dominating step duration and ensures consistent progress across the cluster, even under packet loss or transient congestion.

% \textcolor{blue}{
Since \name{} omits transport-layer recovery, it relies on the ML model's inherent tolerance to partial data. 
Critical information (like activation shards) can be prioritized and split across packets for partial recovery, with lightweight coding schemes (\eg, XOR, Hadamard) used to reconstruct lost fragments---mitigating the impact on model accuracy.
% }
%\ertza{Response point 1}
%\textcolor{orange}{
%As \name{} omits transport-layer recovery, it relies on the ML model’s inherent tolerance to operate with partial data. 
%Building on this tolerance, \name{} applies priority-aware handling at the transport level: critical activation shards and routing metadata can be fragmented and dispersed across multiple packets to enable partial recovery even under repeated loss, while lightweight decoding and reconstruction (\eg, XOR or Hadamard schemes) can restore missing fragments to bound their impact on model accuracy.
%}

By shifting transport semantics to match ML's needs, \name{} enables scalable communication that is robust to loss, simple to implement, and optimized for tail latency.

\vspace{-8pt}
\section{Preliminary Results}
\label{sec:evaluation}

We evaluate \name{} to validate our core claim: simplifying the RDMA transport layer for ML workloads improves tail latency, reduces hardware overhead, and enhances system resilience. 
Our evaluation spans three axes---latency, resource efficiency, and fault tolerance---using both FPGA prototypes and cluster-scale simulation.

\begin{figure}[t]
\centering
\includegraphics[width=0.9\linewidth]{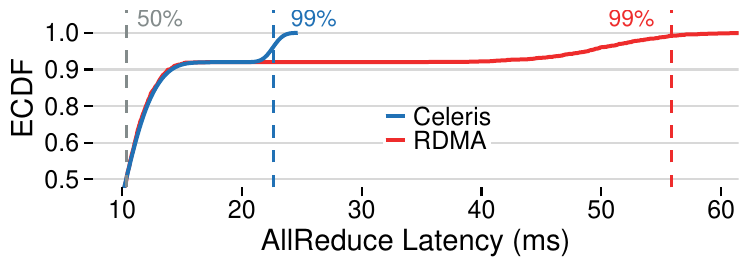}
\vspace{-8pt}
\caption{\bf AllReduce step times under background contention. \name{} bounds tail latency with per-step timeouts.}
\label{fig:ecdf-comparison}
\vspace{-17pt}
\end{figure}

\vspace{2pt}\noindent
{\em Evaluation Setup.} We implement \name{} on an AMD Alveo U250 FPGA using Coyote, an open-source RoCE-compatible NIC shell. 
To support RoCE, IRN~\cite{mittal2018revisiting}, and SRNIC~\cite{wang2023srnic} baselines, we extend Coyote's QP context to 407B, 596B, and 210B, respectively (Table~\ref{tab:rdma-evolution}). 
In contrast, \name{} uses a small 52B per-QP by eliminating retransmissions, in-order delivery, and packet tracking. 
Packets carry an explicit offset, enabling direct placement without reorder buffers.

For cluster-scale simulation, we integrate \name{} with AstraSim and NS-3, modeling a 128-node Clos network.
Each node performs 25MB rounds while randomized, bursty background traffic is injected to create contention.

%\vspace{-10pt}
%\subsection{Performance: Reduction in Tail Latency}
\vspace{4pt}
\noindent{\em A. Performance: Reduction in Tail Latency
\vspace{3pt}\\}
We evaluate tail latency under contention using the above 128-node AllReduce setup.
The baseline uses a RoCE-style RDMA stack with retransmissions and in-order delivery.
As shown in Figure~\ref{fig:ecdf-comparison}, it suffers from high tail latency, with the 99th percentile exceeding 5$\times$ the median due to retransmission delays and head-of-line blocking.

\name{} mitigates this by eliminating transport-layer recovery and instead applying a software-managed timeout per collective step. 
Once the timeout expires, each node finalizes the round using data received up to that point, discarding any late packets. 
We set the timeout to the \emph{median plus one standard deviation} of the baseline distribution. 
Figure~\ref{fig:ecdf-comparison} shows that \name{} cuts the 99th-percentile latency by 2.3$\times$ while preserving median latency. 
Even without retransmissions, less than 1\% of total data is lost, as most nodes complete transmission before the timeout. 
As discussed in \S\ref{sec:background}, this level of loss is well within ML's convergence tolerance.

%\vspace{-10pt}
%\subsection{Resource Efficiency: Reduced NIC Logic and Memory}
\vspace{4pt}
\noindent{\em B. Resource Efficiency: Reduced NIC Logic and Memory
\vspace{3pt}\\}
\name{} reduces hardware complexity by eliminating stateful transport-layer components such as retransmission queues, sequence tracking, and reorder logic. 
These modules consume significant logic and memory in traditional RDMA designs, especially when supporting thousands of QPs. 
We evaluate this by synthesizing \name{} and three baselines---RoCE, IRN, and SRNIC---using Vivado 2022.1 on an AMD Alveo U250 FPGA with 10K QPs. 
As shown in Table~\ref{tab:fpga-resource}, \name{} reduces LUTs by up to 6.6\%, LUTRAMs by 10.2\%, and FFs by 5.2\%. BRAM usage drops by 63.5--72.7\% compared to RoCE and IRN, due to the elimination of bitmaps, per-QP window state, and reassembly buffers. 
Power also improves by up to 9\%, reflecting reduced switching activity across datapath FSMs.

To assess ASIC feasibility, we apply standard FPGA-to-ASIC scaling models for 7nm TSMC technology. 
\name{} has the smallest area, using approximately 57\% less silicon than IRN and about 28\% less than SRNIC.
This compact footprint improves integration density and lowers thermal and validation complexity. 
By reducing the NIC to its essential functions---DMA, header parsing, and congestion control---\name{} not only improves performance but also delivers a streamlined, scalable transport engine for constrained environments.

% \begin{table}[t]
%   \centering
%   \footnotesize
%   \begin{tabular}{lrrrr>{\columncolor[HTML]{cfe2f3}}r}
%   \toprule
%   \textbf{Metric} & \textbf{RoCE} & \textbf{IRN} & \textbf{SRNIC} & \textbf{UCCL} & \textbf{\name{}} \\
%   \midrule
%   LUT & 312{,}449 & 319{,}567 & 304{,}497 & 311{,}244 & 298{,}435 \\
%   LUTRAM & 23{,}277 & 24{,}221 & 22{,}460 & 23{,}179 & 21{,}743 \\
%   FF & 562{,}129 & 573{,}116 & 551{,}526 & 560{,}579 & 542{,}972 \\
%   BRAM & 1450.5 & 1941.5 & 939.5 & 1347 & 529.5 \\
%   Power (W) & 34.7 & 35.9 & 33.5 & 34.5 & 32.5 \\
%   MTBF (hrs) & 42.8 & 34.3 & 57.8 & 45.2 & 80.5 \\
%   \bottomrule
%   \end{tabular}
%   \vspace{-5pt}
%   \caption{\bf \name{} delivers the best balance of area, power, and reliability across FPGA/ASIC metrics.}
%   \label{tab:fpga-resource}
%   \vspace{-18pt}
% \end{table}

\begin{table}[t]
  \centering
  \footnotesize
  \begin{tabular}{lrrrr>{\columncolor[HTML]{cfe2f3}}r}
  \toprule
  \textbf{Metric} & \textbf{RoCE} & \textbf{IRN} & \textbf{SRNIC} & \textbf{\name{}} \\
  \midrule
  LUT & 312{,}449 & 319{,}567 & 304{,}497 & 298{,}435 \\
  LUTRAM & 23{,}277 & 24{,}221 & 22{,}460 & 21{,}743 \\
  FF & 562{,}129 & 573{,}116 & 551{,}526  & 542{,}972 \\
  BRAM & 1450.5 & 1941.5 & 939.5 & 529.5 \\
  Power (W) & 34.7 & 35.9 & 33.5 & 32.5 \\
  MTBF (hrs) & 42.8 & 34.3 & 57.8 & 80.5 \\
  \bottomrule
  \end{tabular}
  \vspace{-5pt}
  \caption{\bf \name{} delivers the best balance of area, power, and reliability across FPGA/ASIC metrics.}
  \label{tab:fpga-resource}
  \vspace{-18pt}
\end{table}

%\vspace{-10pt}
%\subsection{Resilience: Improved Hardware Fault Tolerance}
\vspace{4pt}
\noindent{\em C. Resilience: Improved Hardware Fault Tolerance
\vspace{3pt}\\}
Transport-layer reliability features such as retry engines, sequence tracking, and reorder queues introduce not only complexity but also vulnerability. 
These mechanisms rely on SRAM-backed per-QP state and tightly coupled control logic, which are known sources of soft errors in high-density FPGA and NIC deployments. 
To quantify this, we use the Xilinx SEU Estimator~\cite{amd-seu-estimator} to model Mean Time Between Failures (MTBF) for a 15,000-node datacenter operating at 100\textdegree C. 
We assume a 10\% CRAM essential bit ratio and a standard SRAM FIT rate of $10^{-11}$ per bit~\cite{amd-seu-estimator}. 
Under this model, RoCE and IRN---each maintaining hundreds of bytes of control state per QP---show MTBFs of 42.8 and 34.3 hours, respectively (Table~\ref{tab:fpga-resource}). 
SRNIC improves this to 57.8 hours by offloading some recovery logic to software. 
In contrast, \name{}, with just 52B of per-QP state and no transport-layer recovery, achieves an MTBF of 80.5 hours (nearly 2$\times$ improvement).

\vspace{-8pt}
\section{Conclusion}
\label{sec:conclusion}

\name{} rethinks RDMA transport for ML workloads by eliminating delivery guarantees in favor of tail-optimal, loss-tolerant performance.
By simplifying the datapath and aligning with ML's resilience to loss and partial data, it offers a scalable, efficient, and more robust foundation for collective communication at the cluster scale.
Our early prototype reduces 99th-percentile latency by 2.3$\times$, cuts BRAM usage by over 70\%, and nearly doubles hardware MTBF.
These gains come not from additional complexity, but from removing complex reliability mechanisms that ML does not need.
\name{} shows that ML-aware transport can deliver both simplicity and performance at scale.

{\footnotesize
\vspace{-10pt}
\bibliographystyle{IEEEtran}
\bibliography{paper}
}

\end{document}